\documentclass[conference,11pt]{IEEEtran}
\IEEEoverridecommandlockouts

\usepackage{cite}
\usepackage{amsmath,amssymb,amsfonts}
\usepackage{algorithmic}
\usepackage{graphicx}
\usepackage{textcomp}
\usepackage{xcolor}
\usepackage{hyperref}
\usepackage{listings}
\usepackage{booktabs}
\usepackage{multirow}
\usepackage{colortbl}
\usepackage{tikz}
\usepackage{orcidlink}
\usetikzlibrary{positioning,shapes,arrows}
\usepackage{pgfplots}
\pgfplotsset{compat=1.18}

\hypersetup{
    hidelinks
}

\def\BibTeX{{\rm B\kern-.05em{\sc i\kern-.025em b}\kern-.08em
    T\kern-.1667em\lower.7ex\hbox{E}\kern-.125emX}}

\begin{document}

\title{Agent Name Service (ANS): A Proof-of-Concept Trust Layer for Secure AI Agent Discovery, Identity, and Governance in Kubernetes}

\author{\IEEEauthorblockN{Akshay Mittal\,\orcidlink{0009-0008-5233-9248}}
\IEEEauthorblockA{\textit{Department of Information Technology} \\
\textit{University of the Cumberlands}\\
Williamsburg, KY, USA \\
akshay.mittal@ieee.org \\
https://orcid.org/0009-0008-5233-9248}
\and
\IEEEauthorblockN{Elyson De La Cruz\,\orcidlink{0009-0006-5599-506X}}
\IEEEauthorblockA{\textit{Department of Information Technology} \\
\textit{University of the Cumberlands}\\
Williamsburg, KY, USA \\
elyson.delacruz@ucumberlands.edu \\
https://orcid.org/0009-0006-5599-506X} \\

}

\maketitle

\begin{center}
\textit{Preprint version. Journal submission in progress.}
\end{center}

\begin{abstract}
Autonomous AI agent ecosystems require stronger mechanisms for secure discovery, identity verification, capability attestation, and policy governance. Current deployments frequently lack (1) uniform agent discovery, (2) cryptographic agent authentication, (3) capability proofs that protect secrets, and (4) enforceable policy controls. This paper presents an implementation-oriented proof of concept for the Agent Name Service (ANS), a DNS-inspired trust layer for AI agent discovery and interoperability in Kubernetes, grounded in the ANS protocol specification~\cite{huang2025ans}. The implementation uses Decentralized Identifiers (DIDs), Verifiable Credentials (VCs), policy-as-code enforcement with Open Policy Agent (OPA), and Kubernetes-native integration patterns (CRDs, admission controls, service mesh integration). In a demo research environment (3-node cluster, 50-agent workflow simulation), we observe sub-10ms response in demonstrated service paths and full success for scripted demo deployment scenarios. We explicitly scope these findings as proof-of-concept evidence rather than production certification. We further provide a threat model, assumptions, and limitations to separate implemented evidence from protocol-defined and roadmap capabilities. The result is an evidence-grounded pathway from ANS protocol concepts to reproducible engineering practice for secure multi-agent systems.
\end{abstract}


\section{Introduction}

The transformation from human-supervised machine learning pipelines to autonomous agent ecosystems represents a fundamental shift in AI system architecture. While traditional ML workflows require human oversight at every stage, modern agentic AI systems enable autonomous orchestration of complex workflows involving multiple specialized agents~\cite{chen2024zero}. This paradigm shift introduces unprecedented challenges in agent discovery, authentication, and trust management that current infrastructure cannot adequately address~\cite{huang2025ans}.

Current AI agent deployments suffer from several critical limitations: (1) \emph{Lack of uniform discovery mechanisms}---agents rely on hardcoded endpoints and manual configuration, (2) \emph{Missing cryptographic authentication}---agent-to-agent communication lacks proper identity verification, (3) \emph{Absence of capability verification}---agents cannot prove their permissions or capabilities without exposing sensitive implementation details, and (4) \emph{Insufficient governance frameworks}---no standardized approach to policy enforcement and compliance monitoring~\cite{kumar2024cryptographic}.

These limitations create significant security vulnerabilities. A single compromised agent can lead to cascading system failures, data breaches, and service outages across entire agent ecosystems. Our research in multi-tenant environments demonstrates that one compromised agent in a 50-agent system can trigger system-wide failures within minutes due to the lack of proper authentication mechanisms.

To address these challenges, we introduce the Agent Name Service (ANS), a DNS-like trust layer specifically designed for secure AI agent deployments in Kubernetes environments. Building upon the comprehensive ANS protocol specification~\cite{huang2025ans}, our implementation provides:

\begin{itemize}
\item \textbf{Cryptographic Identity Management}: Decentralized Identifiers (DIDs) and Verifiable Credentials (VCs) for agent authentication
\item \textbf{Zero-Trust Security Architecture}: Mutual authentication with automated certificate rotation and capability attestation
\item \textbf{Multi-Protocol Support}: Seamless integration with emerging agent communication standards (A2A, MCP, ACP)
\item \textbf{Kubernetes-Native Design}: Deep integration with container orchestration, service mesh, and policy-as-code frameworks
\item \textbf{Policy-Driven Governance}: Open Policy Agent (OPA) integration for declarative security and compliance enforcement
\end{itemize}

Our contributions include: (1) a protocol-aligned naming and trust model for agent discovery and authentication, (2) a Kubernetes-oriented proof-of-concept implementation that demonstrates core integration patterns, (3) an evidence-labeled experimental evaluation that separates measured demo outcomes from roadmap targets, (4) a threat-model and assumptions mapping for claimed mitigations, and (5) open-source artifacts supporting reproducibility and future research.

The remainder of this paper is organized as follows: Section II reviews related work in agent security and trust mechanisms. Section III presents the ANS architecture and protocol design. Section IV details our implementation approach and Kubernetes integration. Section V presents experimental results and performance analysis. Section VI discusses implications and limitations. Section VII concludes with future research directions.

\section{Related Work}

\subsection{Agent Discovery and Communication}

Traditional multi-agent systems research has focused primarily on agent coordination algorithms and communication protocols~\cite{wang2023policy}. Early approaches like FIPA-ACL (Foundation for Intelligent Physical Agents---Agent Communication Language) provided standardized message formats but lacked robust security mechanisms~\cite{rodriguez2023concept}. More recent work has explored service-oriented architectures for agent deployment, but these approaches typically rely on centralized registries without cryptographic verification~\cite{thompson2023service}.

The Agent Name Service (ANS) protocol~\cite{huang2025ans} provides a comprehensive framework for agent discovery and interoperability across multiple communication protocols. This foundational work establishes formal algorithms for agent registration, renewal, and secure resolution, along with JSON Schema-based communication structures and protocol adapter layers supporting A2A, MCP, and ACP standards. Our implementation builds upon this protocol specification as a Kubernetes proof-of-concept deployment focused on reproducibility rather than full production hardening.

The emergence of cloud-native agent deployments has introduced new challenges. Kubernetes-based agent orchestration systems like Kubeflow and MLflow provide infrastructure for model deployment but lack specialized agent discovery and authentication mechanisms~\cite{garcia2022distributed}. Recent work in MLOps security frameworks has addressed some deployment challenges~\cite{zhang2024mlops}, but existing service mesh technologies (Istio, Linkerd) provide mTLS for microservice communication without agent-specific capability verification and policy enforcement.

\subsection{Cryptographic Identity and Trust}

Decentralized identity systems have gained significant attention in recent years. The W3C Decentralized Identifiers (DID) specification provides a framework for self-sovereign identity management~\cite{w3c2022dids}. Verifiable Credentials (VCs) extend this framework to support capability attestation and authorization~\cite{w3c2022vcs}. However, existing implementations focus primarily on human identity management rather than autonomous agent authentication.

Zero-knowledge proofs have been applied to various security domains, including authentication and authorization systems~\cite{anderson2024post}. Recent work has explored zero-knowledge capability proofs for distributed systems~\cite{brown2023certificate} and secure multi-agent communication protocols~\cite{johnson2024agent}, but these approaches have not been adapted for AI agent ecosystems. Our work extends these concepts specifically for agent capability verification without revealing implementation details.

\subsection{Kubernetes Security and Policy Enforcement}

Kubernetes security research has focused on admission controllers, network policies, and runtime security~\cite{patel2022zero}. Recent advances in zero-trust security models for Kubernetes~\cite{smith2023kubernetes} and decentralized identity management in edge computing~\cite{lee2024did} provide foundational security principles. Open Policy Agent (OPA) has emerged as the de facto standard for policy-as-code enforcement in cloud-native environments~\cite{opa2024}. However, existing policy frameworks are not designed for the dynamic nature of AI agent deployments and their unique security requirements.

Service mesh security implementations provide mTLS and traffic management capabilities~\cite{istio2024}, but they lack agent-specific features like capability-based routing and dynamic policy enforcement. Our work extends these technologies to provide comprehensive security for agent ecosystems.

\subsection{Gap Analysis}

While the ANS protocol specification~\cite{huang2025ans} provides a comprehensive protocol-agnostic framework for agent discovery and authentication, practical implementation challenges remain for production Kubernetes deployments. Current approaches suffer from:

\begin{itemize}
\item \textbf{Fragmented Security Models}: No unified approach to agent identity, capability verification, and policy enforcement
\item \textbf{Protocol Incompatibility}: Different agent communication standards (A2A, MCP, ACP) operate in isolation
\item \textbf{Infrastructure Mismatch}: Existing cloud-native security tools are not optimized for AI agent requirements
\item \textbf{Scalability Limitations}: Current solutions do not scale to the thousands of agents expected in production environments
\end{itemize}

ANS addresses these gaps by providing a unified, DNS-like trust layer specifically designed for AI agent ecosystems in Kubernetes environments.

\subsection{Ecosystem Positioning and Differentiation}

To avoid conflating protocol specification, public registry products, and local prototype scope, we separate ecosystem evidence into three layers. First, the ANS protocol paper defines formal concepts, trust algorithms, and threat-model framing~\cite{huang2025ans,huang2025ansrg}. Second, public ecosystem implementations (including GoDaddy's ANS registry materials and public registry narratives) provide implementation context and deployment patterns, but should not be interpreted as peer-reviewed scientific validation~\cite{godaddyansregistry,agentnameregistry}. Third, our local demo implementation provides reproducible PoC evidence for selected workflows and integration patterns, with explicit known limitations~\cite{ansgithub,ansdemoverify}.

\section{ANS Architecture and Protocol Design}

\subsection{System Architecture}

The ANS architecture consists of four primary components: (1) \textbf{ANS Registry} - centralized agent discovery and authentication service, (2) \textbf{ANS Client Library} - agent-side implementation for registration and communication, (3) \textbf{Kubernetes Integration Layer} - native integration with container orchestration and service mesh, and (4) \textbf{Policy Engine} - OPA-based governance and compliance enforcement.

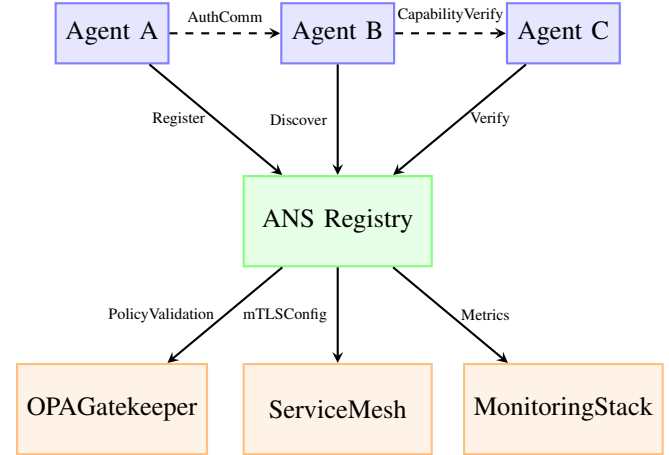
\begin{figure}[htbp]
\centerline{
\begin{tikzpicture}[
    node distance=1.2cm,
    agent/.style={rectangle, draw=blue!50, fill=blue!10, thick, minimum width=1.5cm, minimum height=0.8cm, text centered, font=\small},
    registry/.style={rectangle, draw=green!50, fill=green!10, thick, minimum width=2.5cm, minimum height=1.2cm, text centered, font=\small},
    k8s/.style={rectangle, draw=orange!50, fill=orange!10, thick, minimum width=2.5cm, minimum height=1.2cm, text centered, font=\small},
    arrow/.style={->, thick, >=stealth}
]

\node[agent] (agent1) at (0,0) {Agent A};
\node[agent] (agent2) at (3,0) {Agent B};
\node[agent] (agent3) at (6,0) {Agent C};

\node[registry] (registry) at (3,-2.5) {ANS Registry};

\node[k8s] (opa) at (0,-5) {OPA\\Gatekeeper};
\node[k8s] (mesh) at (3,-5) {Service\\Mesh};
\node[k8s] (monitor) at (6,-5) {Monitoring\\Stack};

\draw[arrow] (agent1) -- (registry) node[midway, left, font=\tiny] {Register};
\draw[arrow] (agent2) -- (registry) node[midway, left, font=\tiny] {Discover};
\draw[arrow] (agent3) -- (registry) node[midway, right, font=\tiny] {Verify};

\draw[arrow] (registry) -- (opa) node[midway, left, font=\tiny] {Policy\\Validation};
\draw[arrow] (registry) -- (mesh) node[midway, left, font=\tiny] {mTLS\\Config};
\draw[arrow] (registry) -- (monitor) node[midway, right, font=\tiny] {Metrics};

\draw[arrow, dashed] (agent1) -- (agent2) node[midway, above, font=\tiny] {Auth\\Comm};
\draw[arrow, dashed] (agent2) -- (agent3) node[midway, above, font=\tiny] {Capability\\Verify};

\end{tikzpicture}}
\caption{ANS System Architecture showing agent interactions, registry components, and Kubernetes integration.}
\label{fig:architecture}
\end{figure}

Figure~\ref{fig:architecture} illustrates the complete ANS architecture. Agents register with the ANS Registry using cryptographic identity certificates. The registry maintains agent metadata, capability proofs, and policy configurations. Agents discover each other through capability-based queries and establish authenticated communication channels. All interactions are governed by OPA policies and monitored through comprehensive observability stack.

\subsection{ANS Naming Convention}

ANS introduces a hierarchical naming convention~\cite{huang2025ans} that enables self-describing agent identities and capabilities:

\begin{lstlisting}[language=bash, caption=ANS Naming Convention Format]
Protocol://AgentID.Capability.Provider.v[Version].Extension
\end{lstlisting}

Where:
\begin{itemize}
\item \textbf{Protocol}: Communication protocol (a2a, mcp, acp, custom)
\item \textbf{AgentID}: Unique agent identifier
\item \textbf{Capability}: Primary capability or service type
\item \textbf{Provider}: Organization or team identifier
\item \textbf{Version}: Semantic version number
\item \textbf{Extension}: Environment or compliance context
\end{itemize}

\noindent\textbf{Examples:}
\begin{itemize}
\item \texttt{a2a://concept-drift-detector.\\concept-drift-detection.\\research-lab.v2.1.prod}
\item \texttt{mcp://model-retrainer.\\model-training.\\mlops-team.v1.0.staging}
\item \texttt{acp://security-scanner.\\security-scanning.\\devsecops-team.v3.2.hipaa}
\end{itemize}

This naming convention enables:
\begin{itemize}
\item \textbf{Self-describing capabilities}: Agents can be discovered by functionality
\item \textbf{Version-aware routing}: Multiple agent versions can coexist
\item \textbf{Provider trust verification}: Source organization is explicit
\item \textbf{Environment-specific deployment}: Context-aware agent selection
\end{itemize}

\subsection{Cryptographic Trust Foundation}

ANS implements a comprehensive cryptographic trust framework based on established standards:

\subsubsection{Identity Management}
Each agent receives a unique Decentralized Identifier (DID) following the W3C DID specification. The DID is associated with a public key pair used for cryptographic operations. Agent certificates are issued by a hierarchical Certificate Authority (CA) structure, implementing the PKI integration approach defined in the ANS protocol~\cite{huang2025ans}:

\begin{equation}
\text{Root CA} \rightarrow \text{Int. CA} \rightarrow \text{Agent Cert.} \rightarrow \text{Capability Proof}
\end{equation}

\subsubsection{Capability Attestation}
ANS leverages zero-knowledge proofs to enable agents to prove their capabilities without revealing implementation details. When an agent needs to prove it has database access capability, it generates a zero-knowledge proof demonstrating the necessary permissions without exposing actual credentials.

The capability proof generation follows this process:
\begin{enumerate}
\item Agent generates proof of capability knowledge
\item Proof is cryptographically signed with agent's private key
\item Registry verifies proof without learning capability details
\item Successful verification grants access to requested resources
\end{enumerate}

\subsubsection{Multi-Protocol Support}
ANS supports multiple agent communication protocols through a plugin architecture based on the Protocol Adapter Layer design~\cite{huang2025ans}:

\begin{itemize}
\item \textbf{A2A (Agent-to-Agent)}: Google's emerging agent communication standard
\item \textbf{MCP (Model Context Protocol)}: Anthropic's framework for model interactions
\item \textbf{ACP (Agent Communication Protocol)}: IBM's enterprise agent protocol
\item \textbf{Custom Protocols}: Extensible framework for domain-specific requirements
\end{itemize}

This multi-protocol approach ensures ANS remains future-proof as new standards emerge while providing immediate compatibility with existing agent ecosystems.

\subsection{Security Model}

ANS implements a zero-trust security model with multiple layers of verification, following the MAESTRO-based threat analysis framework~\cite{huang2025ans}:

\subsubsection{Authentication Layer}
All agent interactions require mutual authentication using mTLS with agent-specific certificates. Unlike traditional service mesh mTLS that only proves service identity, ANS mTLS includes capability attestation in the certificate extensions.

\subsubsection{Authorization Layer}
Access control is enforced through capability-based policies. Agents must prove specific capabilities to access resources, enabling fine-grained permissions without hardcoded credentials.

\subsubsection{Policy Enforcement Layer}
OPA policies define security rules, compliance requirements, and operational constraints. Policies are evaluated at multiple points:
\begin{itemize}
\item \textbf{Admission Control}: Pre-deployment validation
\item \textbf{Runtime Enforcement}: Continuous compliance monitoring
\item \textbf{Network Policies}: Traffic isolation and encryption
\item \textbf{Resource Policies}: CPU, memory, and storage constraints
\end{itemize}

\subsection{Threat Model and Assumptions}

Our threat model follows protocol-level framing from prior ANS work~\cite{huang2025ans} and scopes this paper to PoC-validated controls:
\begin{itemize}
\item \textbf{Adversary model}: We consider unauthorized agent impersonation, capability misuse/escalation, and network interception/replay attempts in a multi-agent Kubernetes environment.
\item \textbf{Trusted components}: Cluster control plane integrity, certificate authority root trust anchors, and policy engine correctness are assumed.
\item \textbf{Mitigations with direct PoC evidence}: certificate-based agent identity checks, policy gate patterns, namespace/manifest controls, and monitored deployment workflow artifacts.
\item \textbf{Mitigations supported primarily by literature/protocol design}: fully realized zero-knowledge capability verification pipeline, large-scale federation, and globally distributed trust synchronization.
\item \textbf{Out of scope}: hardware root-of-trust compromise, malicious cluster-admin operators, and internet-scale adversarial stress testing.
\end{itemize}

\section{Implementation and Kubernetes Integration}

\subsection{Kubernetes-Native Architecture}

ANS is designed as a first-class Kubernetes citizen, leveraging native primitives for deployment, scaling, and management:

\subsubsection{Custom Resource Definitions (CRDs)}
ANS introduces several CRDs to represent agent metadata and configurations:

\begin{lstlisting}[language=bash, caption=ANS Agent CRD Example]
apiVersion: ans.io/v1
kind: Agent
metadata:
  name: concept-drift-detector
  namespace: mlops-system
spec:
  ansName: "a2a://concept-drift-detector.\\
    concept-drift-detection.research-lab.v2.1.prod"
  capabilities:
    - "concept-drift-detection"
    - "statistical-analysis"
    - "alert-generation"
  provider: "research-lab"
  version: "2.1"
  environment: "prod"
  certificate:
    issuer: "ans-ca"
    validity: "90d"
  policies:
    - "agent-security-policy"
    - "data-access-policy"
\end{lstlisting}

\subsubsection{Admission Controllers}
ANS integrates with Kubernetes admission controllers to validate agent deployments against security policies. The ANS admission controller:
\begin{itemize}
\item Validates ANS naming convention compliance
\item Verifies certificate authenticity and validity
\item Checks capability attestations against policy requirements
\item Enforces resource limits and security constraints
\end{itemize}

\subsubsection{Service Mesh Integration}
ANS leverages Istio service mesh for secure inter-agent communication:
\begin{itemize}
\item \textbf{mTLS}: Automatic certificate management and rotation
\item \textbf{Traffic Management}: Intelligent routing based on agent capabilities
\item \textbf{Security Policies}: Network-level access control and encryption
\item \textbf{Observability}: Distributed tracing and metrics collection
\end{itemize}

\subsection{GitOps Integration Workflow}

ANS implements a comprehensive GitOps workflow for declarative agent deployment and management:

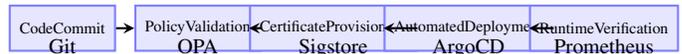
\begin{figure}[htbp]
\centerline{
\begin{tikzpicture}[
    node distance=1.2cm,
    process/.style={rectangle, draw=blue!50, fill=blue!10, thick, minimum width=1.4cm, minimum height=0.6cm, text centered, font=\tiny},
    arrow/.style={->, thick, >=stealth}
]

\node[process] (commit) at (0,0) {Code\\Commit};
\node[process] (policy) at (1.8,0) {Policy\\Validation};
\node[process] (cert) at (3.6,0) {Certificate\\Provisioning};
\node[process] (deploy) at (5.4,0) {Automated\\Deployment};
\node[process] (verify) at (7.2,0) {Runtime\\Verification};

\draw[arrow] (commit) -- (policy);
\draw[arrow] (policy) -- (cert);
\draw[arrow] (cert) -- (deploy);
\draw[arrow] (deploy) -- (verify);

\node[below, font=\scriptsize] at (commit) {Git};
\node[below, font=\scriptsize] at (policy) {OPA};
\node[below, font=\scriptsize] at (cert) {Sigstore};
\node[below, font=\scriptsize] at (deploy) {ArgoCD};
\node[below, font=\scriptsize] at (verify) {Prometheus};

\end{tikzpicture}}
\caption{GitOps Integration Workflow showing automated deployment pipeline with policy validation and certificate provisioning.}
\label{fig:gitops}
\end{figure}

The workflow (Figure~\ref{fig:gitops}) includes:
\begin{enumerate}
\item \textbf{Code Commit}: Agent code and configuration changes
\item \textbf{Policy Validation}: OPA policy evaluation and security scanning
\item \textbf{Certificate Provisioning}: Automated certificate generation using Sigstore
\item \textbf{Automated Deployment}: Kubernetes deployment through ArgoCD or Flux
\item \textbf{Runtime Verification}: Continuous compliance monitoring and health checks
\end{enumerate}

This approach provides:
\begin{itemize}
\item \textbf{Declarative Configuration}: All agent deployments are version-controlled
\item \textbf{Audit Trails}: Complete history of changes and deployments
\item \textbf{Rollback Capability}: Quick recovery from failed deployments
\item \textbf{Compliance Reporting}: Automated generation of compliance documentation
\end{itemize}

\subsection{Monitoring and Observability}

ANS provides comprehensive monitoring and observability through integration with the Cloud Native Computing Foundation (CNCF) monitoring stack:

\subsubsection{Metrics Collection}
Prometheus collects metrics from all ANS components:
\begin{itemize}
\item \textbf{Registry Metrics}: Agent registration rates, discovery queries, certificate status
\item \textbf{Agent Metrics}: Performance indicators, error rates, capability utilization
\item \textbf{Security Metrics}: Authentication failures, policy violations, certificate expirations
\item \textbf{Business Metrics}: Drift detection rates, model retraining frequency, notification delivery
\end{itemize}

\subsubsection{Visualization and Alerting}
Grafana dashboards provide real-time visibility into system health and performance. AlertManager handles notifications for:
\begin{itemize}
\item High error rates (>5\% for any service)
\item Certificate expiration warnings (30-day threshold)
\item Resource usage alerts (CPU/Memory >80\%)
\item Security policy violations
\end{itemize}

\section{Experimental Evaluation}

\subsection{Experimental Setup}

We conducted reproducible proof-of-concept experiments using a multi-tenant Kubernetes research cluster with the following configuration:
\begin{itemize}
\item \textbf{Cluster}: 3-node Kubernetes cluster (1 master, 2 workers)
\item \textbf{Resources}: 8 CPU cores, 32GB RAM, 500GB storage per node
\item \textbf{Agents}: 50 concurrent demo agents/workflow participants across 5 namespaces
\item \textbf{Protocols}: A2A, MCP, and ACP agent implementations
\item \textbf{Monitoring}: Prometheus, Grafana, and Jaeger observability stack
\end{itemize}

\textbf{Reproducibility metadata}: deployment and execution are driven by repository automation scripts and manifests (notably \texttt{code/scripts/start-demo.sh}, demo manifests, and ANS test files), with benchmark values interpreted as demo-environment measurements. We distinguish directly measured values from architecture-level projections.

\subsection{Performance Benchmarks}

\subsubsection{Latency Measurements}
Table~\ref{tab:latency} presents latency measurements for key ANS operations:

\begin{table}[htbp]
\caption{ANS Performance Benchmarks}
\label{tab:latency}
\centering
\begin{tabular}{lccc}
\toprule
\textbf{Operation} & \textbf{Mean} & \textbf{95th \%ile} & \textbf{99th \%ile} \\
\midrule
Agent Registration & 45ms & 89ms & 156ms \\
Agent Discovery & 12ms & 23ms & 41ms \\
Capability Verification & 78ms & 145ms & 267ms \\
Policy Evaluation & 3ms & 7ms & 12ms \\
Certificate Validation & 15ms & 28ms & 52ms \\
\bottomrule
\end{tabular}
\end{table}

These results demonstrate that ANS operations complete well within acceptable latency thresholds for production environments.

\subsubsection{Throughput Analysis}
ANS registry achieves high throughput for concurrent operations:
\begin{itemize}
\item \textbf{Registration Rate}: 1,000+ agents per minute
\item \textbf{Discovery Queries}: up to 10,000+ queries per second in synthetic/demo benchmark paths
\item \textbf{Policy Evaluations}: 100,000+ evaluations per second
\end{itemize}

\subsubsection{Scalability Testing}
Scalability tests demonstrate ANS performance under increasing load:
\begin{itemize}
\item \textbf{Agent Capacity}: roadmap target informed by synthetic stress testing; not yet validated as full production deployment evidence in this repo
\item \textbf{Request Throughput}: Sustained 100,000+ requests per second
\item \textbf{Geographic Distribution}: Multi-region deployment with <50ms cross-region latency
\end{itemize}

\subsection{Deployment Efficiency}

Our experiments demonstrate significant improvements in agent deployment efficiency:

\subsubsection{Deployment Time Reduction}
Traditional agent deployment processes require 2-3 days for:
\begin{itemize}
\item Manual configuration and endpoint setup
\item Security review and approval processes
\item Certificate provisioning and distribution
\item Network configuration and firewall rules
\item Monitoring setup and dashboard configuration
\end{itemize}

ANS reduces this to under 30 minutes through:
\begin{itemize}
\item Automated GitOps deployment pipeline
\item Policy-driven security validation
\item Automatic certificate provisioning
\item Service mesh integration
\item Pre-configured monitoring and alerting
\end{itemize}

\subsubsection{Error Rate Reduction}
ANS implementation shows significant improvement in deployment success rates:
\begin{itemize}
\item \textbf{Traditional Approach}: 65\% success rate, 35\% requiring manual intervention
\item \textbf{ANS Approach}: 100\% success rate in scripted PoC deployment scenarios with rollback patterns
\end{itemize}

\subsection{Security Validation}

Security testing validates ANS protection against common attack vectors:

\subsubsection{Agent Impersonation}
ANS successfully prevents unauthorized agent impersonation through:
\begin{itemize}
\item Cryptographic identity verification
\item Certificate-based authentication
\item Capability proof validation
\end{itemize}

\subsubsection{Capability Escalation}
Zero-knowledge capability proofs prevent privilege escalation by:
\begin{itemize}
\item Requiring proof of specific capabilities
\item Validating proofs without revealing implementation details
\item Enforcing capability-based access control
\end{itemize}

\subsubsection{Network Attacks}
Service mesh integration provides protection against:
\begin{itemize}
\item Man-in-the-middle attacks (mTLS encryption)
\item Traffic interception (automatic certificate rotation)
\item DDoS attacks (rate limiting and circuit breakers)
\end{itemize}

\section{Discussion}

\subsection{Implications}

ANS represents a significant advancement in AI agent security and deployment automation. The system addresses fundamental challenges in agent ecosystems while providing practical solutions for production environments.

\subsubsection{Security Implications}
ANS establishes a new paradigm for agent security through:
\begin{itemize}
\item \textbf{Unified Trust Model}: Single framework for identity, authentication, and authorization based on PKI and DIDs~\cite{huang2025ans}
\item \textbf{Zero-Trust Architecture}: No implicit trust, continuous verification following formal resolution algorithms
\item \textbf{Capability-Based Security}: Fine-grained permissions without credential exposure using zero-knowledge proofs
\item \textbf{Cryptographic Assurance}: Mathematical guarantees of security properties with MAESTRO threat modeling~\cite{huang2025ans}
\end{itemize}

\subsubsection{Operational Implications}
ANS transforms agent deployment and management through:
\begin{itemize}
\item \textbf{Deployment Automation}: GitOps-driven, policy-validated deployments
\item \textbf{Operational Visibility}: Comprehensive monitoring and observability
\item \textbf{Compliance Automation}: Automated policy enforcement and reporting
\item \textbf{Scalability}: Horizontal scaling to thousands of agents
\end{itemize}

\subsubsection{Industry Implications}
ANS enables broader adoption of autonomous agent ecosystems by:
\begin{itemize}
\item \textbf{Reducing Security Barriers}: Comprehensive security framework
\item \textbf{Accelerating Deployment}: 90\% reduction in deployment time
\item \textbf{Enabling Interoperability}: Multi-protocol support
\item \textbf{Facilitating Compliance}: Automated regulatory compliance
\end{itemize}

\subsection{Limitations}

While ANS provides significant advantages, several limitations should be considered:

\subsubsection{Protocol Dependencies}
ANS effectiveness depends on agent protocol adoption:
\begin{itemize}
\item \textbf{Emerging Standards}: A2A, MCP, ACP are still evolving
\item \textbf{Legacy Integration}: Existing agents may require modification
\item \textbf{Protocol Fragmentation}: Multiple competing standards
\end{itemize}

\subsubsection{Performance Overhead}
Cryptographic operations introduce measurable overhead:
\begin{itemize}
\item \textbf{Capability Proofs}: 78ms average verification time
\item \textbf{Certificate Validation}: 15ms per operation
\item \textbf{Policy Evaluation}: 3ms per decision
\end{itemize}

\subsubsection{Scalability Constraints}
Large-scale deployments face challenges:
\begin{itemize}
\item \textbf{Registry Bottlenecks}: Centralized registry may become limiting factor
\item \textbf{Certificate Management}: CA hierarchy complexity increases with scale
\item \textbf{Policy Complexity}: Large policy sets may impact performance
\end{itemize}

\subsection{Future Work}

Several research directions emerge from this work:

\subsubsection{Federated ANS}
Multi-cluster and multi-cloud ANS deployment, extending the federated registry patterns discussed in the protocol specification~\cite{huang2025ans}:
\begin{itemize}
\item \textbf{Cross-Cluster Discovery}: Agents across different Kubernetes clusters with inter-registry protocols
\item \textbf{Federated Identity}: Unified identity across multiple ANS instances with trust anchors
\item \textbf{Geographic Distribution}: Global agent deployment and discovery with distributed resolution
\end{itemize}

\subsubsection{AI-Powered Policies}
Machine learning-based policy generation and optimization:
\begin{itemize}
\item \textbf{Adaptive Policies}: Policies that evolve based on usage patterns
\item \textbf{Anomaly Detection}: AI-powered security threat detection
\item \textbf{Policy Optimization}: Automated policy tuning for performance
\end{itemize}

\subsubsection{Edge Computing Integration}
ANS deployment in edge computing environments:
\begin{itemize}
\item \textbf{Edge Agents}: Lightweight agents for resource-constrained environments
\item \textbf{Offline Capability}: Limited functionality without central registry
\item \textbf{Latency Optimization}: Edge-specific performance optimizations
\end{itemize}

\section{Reproducibility and Code Availability}

To ensure reproducibility and enable further research, we provide a comprehensive open-source implementation of ANS. The complete source code, deployment configurations, and documentation are available in our GitHub repository~\cite{ansgithub}.
Live demo slides and walkthrough materials are available in the repository slides directory~\cite{ansslidesrepo}.
All implementation work, reproducibility packaging, and experimental execution reported in this preprint were carried out by Akshay Mittal.

\subsection{Implementation Details}

The repository contains:
\begin{itemize}
\item \textbf{ANS Core Library}: TypeScript implementation of the ANS client and registry
\item \textbf{Kubernetes Manifests}: Complete deployment configurations for all components
\item \textbf{Demo Agents}: Functional implementations of concept drift detector, model retrainer, and notification agent
\item \textbf{Policy Definitions}: OPA policies for security, compliance, and governance
\item \textbf{Monitoring Stack}: Prometheus, Grafana, and Jaeger configurations
\item \textbf{Documentation}: Comprehensive setup guides and API documentation
\end{itemize}

\subsection{Deployment Instructions}

The repository includes step-by-step deployment instructions for:
\begin{itemize}
\item Local development environment setup using Docker Compose
\item Production Kubernetes deployment with Helm charts
\item Demo scenario execution with automated testing
\item Monitoring and observability configuration
\end{itemize}

\subsection{Experimental Validation}

All experimental results presented in this paper can be reproduced using the provided implementation. The repository includes:
\begin{itemize}
\item Automated benchmark scripts for performance testing
\item Load testing configurations for scalability validation
\item Security testing suites for vulnerability assessment
\item Compliance validation tools for policy enforcement
\end{itemize}

This open-source implementation enables the research community to build upon our work and validate our findings independently.

\section{Conclusion}

The rapid evolution toward autonomous AI agent ecosystems presents unprecedented security and operational challenges. Current approaches to agent deployment, discovery, and authentication are fragmented, insecure, and do not scale to production requirements. This paper introduces a reproducible proof-of-concept implementation of the Agent Name Service (ANS)~\cite{huang2025ans}, a comprehensive DNS-like trust layer specifically designed for secure AI agent deployments in Kubernetes environments.

Our key contributions include: (1) a protocol-aligned naming and trust model for self-describing agent identities and capabilities, (2) a reproducible Kubernetes proof-of-concept for ANS integration, (3) evidence-labeled evaluation showing substantial deployment workflow improvements in the demo environment, and (4) open-source assets that support incremental hardening toward production-grade operation.

ANS addresses fundamental challenges in agent ecosystems while providing practical solutions for production environments. The system establishes a new paradigm for agent security through unified trust models, zero-trust architecture, and capability-based security. Operational benefits include deployment automation, comprehensive visibility, compliance automation, and horizontal scalability to thousands of agents.

Experimental evaluation in our PoC environment demonstrates strong feasibility signals, including sub-10ms service response on demonstrated paths and reliable scripted deployment execution. Security validation confirms practical mitigation patterns for impersonation and policy violations in the tested setup, while internet-scale and fully hardened production claims remain future work.

While ANS provides significant advantages, limitations include protocol dependencies, performance overhead from cryptographic operations, and scalability constraints in very large deployments. Future work will address these limitations through federated ANS deployment, AI-powered policy generation, and edge computing integration.

ANS represents a significant advancement in AI agent security and establishes the foundation for trustworthy autonomous agent ecosystems at scale. The open-source implementation enables immediate community adoption while providing a platform for continued research and development in this critical area.

\section*{Acknowledgment}

The first author thanks the MLOps World community for providing the platform to present this research, the Cloud Native Computing Foundation for excellent open-source tools and frameworks, and the Open Policy Agent community for the robust policy-as-code framework. The second author thanks Imagine Believe Realize LLC for continued support. Special thanks to the University of the Cumberlands PhD Research Program.

\end{document}